%% file: ms.tex
\documentclass[11pt]{emulateapj}
\usepackage{apjfonts}

\input psfig

\def\gax{\mathrel{\raise.3ex\hbox{$>$}\mkern-14mu\lower0.6ex\hbox{$\sim$}}}
\def\lax{\mathrel{\raise.3ex\hbox{$<$}\mkern-14mu\lower0.6ex\hbox{$\sim$}}}
\def\gtorder{\mathrel{\raise.3ex\hbox{$>$}\mkern-14mu
             \lower0.6ex\hbox{$\sim$}}}
\def\ltorder{\mathrel{\raise.3ex\hbox{$<$}\mkern-14mu
             \lower0.6ex\hbox{$\sim$}}}

\def\avgm{\langle M\rangle}
\def\avgmhat{\langle M/M_\odot\rangle}

\def\roptlog{$R_{\lambda,O}=1.3 \times 10^{15}$~cm}  
\def\rxraylog{$R_{1/2,X}=2.3 \times 10^{14}$~cm}       

\begin{document}

\title{The Sizes of the X-ray and Optical Emission Regions of RXJ~1131--1231 }

\author{X. Dai\altaffilmark{1}, C.S.\ Kochanek\altaffilmark{2,3}, G. Chartas\altaffilmark{4}, S.\ Koz{\l}owski\altaffilmark{2}, 
  C.W. Morgan\altaffilmark{5}, G. Garmire\altaffilmark{4}, E. Agol\altaffilmark{6} }
 
\altaffiltext{1}{Department of Astronomy, University of Michigan, 500 Church Street, Ann Arbor MI 48109}
\altaffiltext{2}{Department of Astronomy, The Ohio State University, 140 West 18th Avenue, Columbus OH 43210}
\altaffiltext{3}{Center for Cosmology and Astroparticle Physics, The Ohio State University, 140 West 18th Avenue, Columbus OH 43210}
\altaffiltext{4}{Department of Astronomy and Astrophysics, Pennsylvania State University, University Park, PA 16802 }
\altaffiltext{5}{Department of Physics, United States Naval Academy, 572C Holloway Road, Annapolis, MD 21402}
\altaffiltext{6}{Department of Astronomy, University of Washington, 3910 15$^{th}$ Avenue, Seattle WA 98105}

\begin{abstract}
We use gravitational microlensing of the four images of the $z=0.658$ 
quasar RXJ~1131--1231 to measure the sizes of the optical and X-ray 
emission regions of the quasar.  The (face-on) scale length of the 
optical disk at rest frame 400nm is \roptlog, while the half-light radius of the 
rest frame 0.3--17~keV X-ray emission is \rxraylog.  The formal uncertainties are factors of 
$1.6$ and $2.0$, respectively. With the exception of the
lower limit on the X-ray size, the results are very stable against 
any changes in the priors used in the analysis.
Based on the H$\beta$ line-width,
we estimate that the black hole mass is $M_{1131}\simeq 10^8 M_\odot$, which corresponds to a 
gravitational radius of $r_g \simeq 2 \times 10^{13}$~cm.  Thus, the 
X-ray emission is emerging on scales of $\sim 10 r_g$ and the 400~nm emission on 
scales of $\sim 70 r_g$.  A standard thin disk of this size should be significantly
brighter than observed.  Possible solutions are to have a flatter temperature
profile or to scatter a large fraction of the optical flux on larger scales
after it is emitted.  While our calculations were not optimized to constrain the
dark matter fraction in the lens galaxy, dark matter dominated models
are favored.
With well-sampled optical and X-ray light curves over a broad range of
frequencies there will be no difficulty in extending our analysis
to completely map the structure of the accretion disk as a function
of wavelength.
\end{abstract}


\section{Introduction}
\label{sec:introduction}

A significant problem for theoretical studies of quasars is that we cannot
spatially resolve their emission regions to test 
models (e.g. Blaes 2004).  For example, in this paper we study the gravitational
lens RXJ~1231--1131 (RXJ1131 hereafter), where we observe
four images of a $z_s=0.658$ quasar lensed by a $z_l=0.295$ elliptical 
galaxy (Sluse et al.~2003).  Based on the H$\beta$ line-width from Sluse et al.~(2003), 
and a magnification corrected estimate of the continuum luminosity,
we estimate\footnote{Using the 
Bentz et al. (2006) mass normalizations.  For the Kaspi et al. (2005)
normalization we obtain $M_{1131}=(6.9 \pm 1.6) \times 10^7 M_\odot$,
which is consistent with the earlier estimate of Peng et al. (2006) of 
$6 \times 10^{7} M_\odot$ also using the Kaspi et al. (2005) 
normalizations.  We use the Peng et al. (2006) masses in Morgan et al.
(2009) because we lacked spectra for the full sample of objects.}  
that the black hole mass, $M_{BH}$, is 
$M_{1131}=(1.3 \pm 0.3) \times 10^8 M_\odot$.
This corresponds to a gravitational radius of  
\begin{equation}
    r_g =  { G M_{BH} \over c^2 } \simeq \left(1.9 \times 10^{13}\right) 
     \left[ { M_{BH} \over M_{1131}} \right] \hbox{cm},
    \label{eqn:rgrav}
\end{equation}
that subtends only $0.001 h^{-1}$~micro-arcseconds.

Gravity, however, has provided
us with a natural telescope for studying the structure of quasars through the
microlensing produced by stars in the lens galaxy (see the review
by Wambsganss 2006).   Microlensing has a 
natural outer length scale corresponding to the Einstein radius of the stars,
\begin{equation}
    \langle R_E\rangle =
      D_{OS} \left[ { 4 G \avgm D_{LS} \over c^2 D_{OL} D_{OS}} \right]^{1/2}
      = \left(4.6 \times 10^{16}\right) 
            \left[ { \avgm \over M_\odot } \right]^{1/2} \hbox{cm},
     \label{eqn:re}
\end{equation}
where $\avgm$ is the mean mass of the stars, and the distances $D_{OL}$, $D_{OS}$
and $D_{LS}$ are the angular diameter distances between the observer, lens
and source.  The microlenses also generate caustic lines on which the magnification 
diverges, which means that our gravitational telescope can, for all practical
purposes, resolve arbitrarily small sources.  The size of the source
is encoded in the amplitude of the microlensing variability as
the source, lens, and observer move relative to the caustic patterns -- big
sources have smaller variability amplitudes than small sources.  
The technique can be applied to any emission arising from scales more 
compact than a few $\langle R_E\rangle$.  

If we  model the accretion disk by a thermally radiating thin disk with a temperature 
profile of $T \propto R^{-3/4}$ (Shakura \& Sunyaev 1973)\footnote{In 
our present analysis we can neglect the drop in temperature and
emission near the inner edge of the accretion disk as it has
little effect on the results.}, we can measure the scale $R_\lambda$ defined by
the point where the photon energy equals the disk
temperature, $k T = h c/\lambda_{rest}$, by two routes other than microlensing. First, we can
estimate it from the observed flux at some wavelength.  For example,
at I-band the radius is
\begin{equation}
 R_\lambda \simeq 
   { 2.8 \times 10^{15} \over \sqrt{\cos i}} { D_{OS} \over r_H}
    \left[ {\lambda_{obs} \over \mu\hbox{m}} \right]^{3/2} 10^{-0.2(I-19)}  \hbox{cm}.
   \label{eqn:fsize}
\end{equation}
where $r_H=c/H_0$ is the Hubble radius, and $i$ is the inclination angle of the disk.  Based 
on HST observations (Sluse et al. 2006; Koz{\l}owski et al. 2009), we estimate that the 
magnification-corrected flux is $\hbox{I} \simeq 20.7 \pm 0.1$ mag ($\lambda_{obs}=0.814\mu$m), 
which corresponds to an R-band (400~nm in the quasar rest frame) size
of $R_{\lambda,O} = (3.5 \pm 0.2) \times 10^{14} (\cos i)^{-1/2}$~cm
or about $18 r_g$.   The flux size depends on the mean magnification of the images
as $1/\sqrt{\langle \mu\rangle}$, which can introduce a $\sim 50\%$ systematic 
uncertainty into this size estimate.
Second, thin disk theory predicts that
\begin{eqnarray}
   R_\lambda &= &{ 1 \over \pi^2 } \left[ { 45 \over 16} { \lambda^4 r_g \dot{M} \over  h_p} \right]^{1/3} \\
    &= &(2.5 \times 10^{15}) \left[ { \lambda_{rest} \over \mu m} \right]^{4/3} \left[ { M_{BH} \over M_{1131} }\right]^{2/3} 
     \left[ { L \over \eta L_E } \right]^{1/3} \hbox{cm}, \nonumber
     \label{eqn:rdisk}
\end{eqnarray}   
which implies an R-band disk size $R_{\lambda,O}=1.6 \times 10^{15}$~cm 
($82 r_g$) if the disk
is radiating at the Eddington limit $(L/L_E)=1$ with an efficiency
of $\eta=0.1$.  Note that these two size estimates can be 
reconciled only if $(L/\eta L_E)(M_{BH}/M_{1131})^2 \simeq 0.1 (\cos i)^{-3/2}$,
corresponding to a sub-Eddington accretion rate, 
an overestimated black hole mass, or a problem in the disk model 
since there is no evidence for the 1--2~mags of extinction in the
lens galaxy that would be needed raise the flux
size up to that from thin disk theory  (Eqn.~\ref{eqn:rdisk}). 
Adding the inner disk edge or using a simple relativistic disk
model (Novikov \& Thorne 1973, Page \& Thorne 1974) changes this problem little.

The expected size of the X-ray emitting regions is more
problematic because there is no comparably simple model for
our theoretical expectations.  There is a general consensus
that the X-ray continuum emission is due to unsaturated
inverse Compton scattering of soft photons by hot
electrons in a corona surrounding the inner parts of the
accretion disk (see the review by Reynolds \& Nowak 2003), but
the extent and geometrical configuration of the X-ray emission
region is an open question.  The X-ray continuum from the
corona illuminates the disk to produce Fe K$\alpha$ emission 
lines, whose broad widths indicate that they are generated
close to the inner edge of the accretion disk (e.g.
Fabian et al. 2005).

While there were a number of early attempts at estimating accretion
disk sizes using microlensing (e.g. Wambsganss, Schneider \& Paczy\'nski 1990, 
Rauch \& Blandford 1991, Wyithe et al. 2000b,
Wambsganss et al. 2000, Goicoechea et al. 2003), it is only
in the last few years that it has become possible to make large
numbers of microlensing size estimates.  In particular, Pooley
et al. (2007) argue that the optical sizes estimated from microlensing
must be considerably larger than the optical ``flux'' sizes of
Eqn.~\ref{eqn:fsize}. This was confirmed by Morgan et al. (2009)
in a more detailed analysis that also found that the optical sizes
agree better with the thin disk size estimate (Eqn.~\ref{eqn:rdisk})
than the flux size and have a scaling with black hole mass consistent with 
the $M_{BH}^{2/3}$ scaling for Eddington-limited thin disks.  

Recent
studies have started to examine the temperature dependence of disks
through the scaling of disk size with wavelength (Anguita et al. 2008,
Poindexter et al. 2008, Agol et al. 2009, 
Bate et al. 2009, Floyd et al. 2009, Mosquera et al. 2009). 
Studies of the microlensing of the X-ray emission are more limited,
but indicate that the X-ray emission is much more compact than the
optical (Dai et al. 2003, Pooley et al. 2006, 2007, Kochanek et al. 2006, 
Morgan et al. 2008, Chartas et al. 2009), tracking much closer to the
inner edge of the accretion disk.  In this paper we estimate
the sizes of the optical and X-ray emission regions of RXJ1131 using microlensing. 
In \S2 we describe the data and the analysis method.  In \S3
we discuss the results, their implications and directions for
further research.  We use an $\Omega_0=0.3$ flat cosmological
model with $H_0=100 h$~km~s$^{-1}$~Mpc$^{-1}$ and $h=0.7$.

\section{Data and Analysis \label{sec:data}}

\begin{figure}[t]
\centerline{\psfig{file=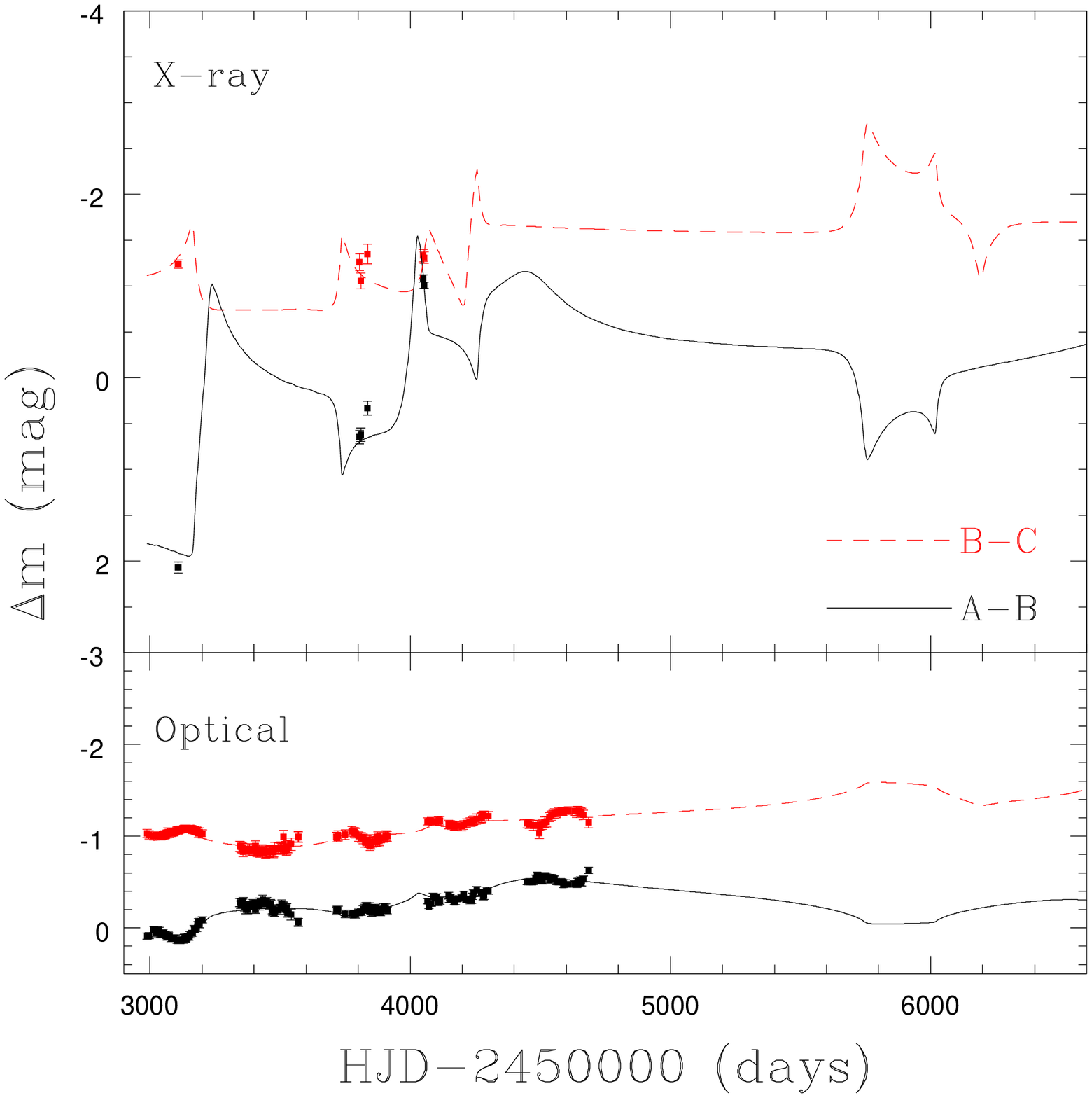,width=3.5in}}
\centerline{\psfig{file=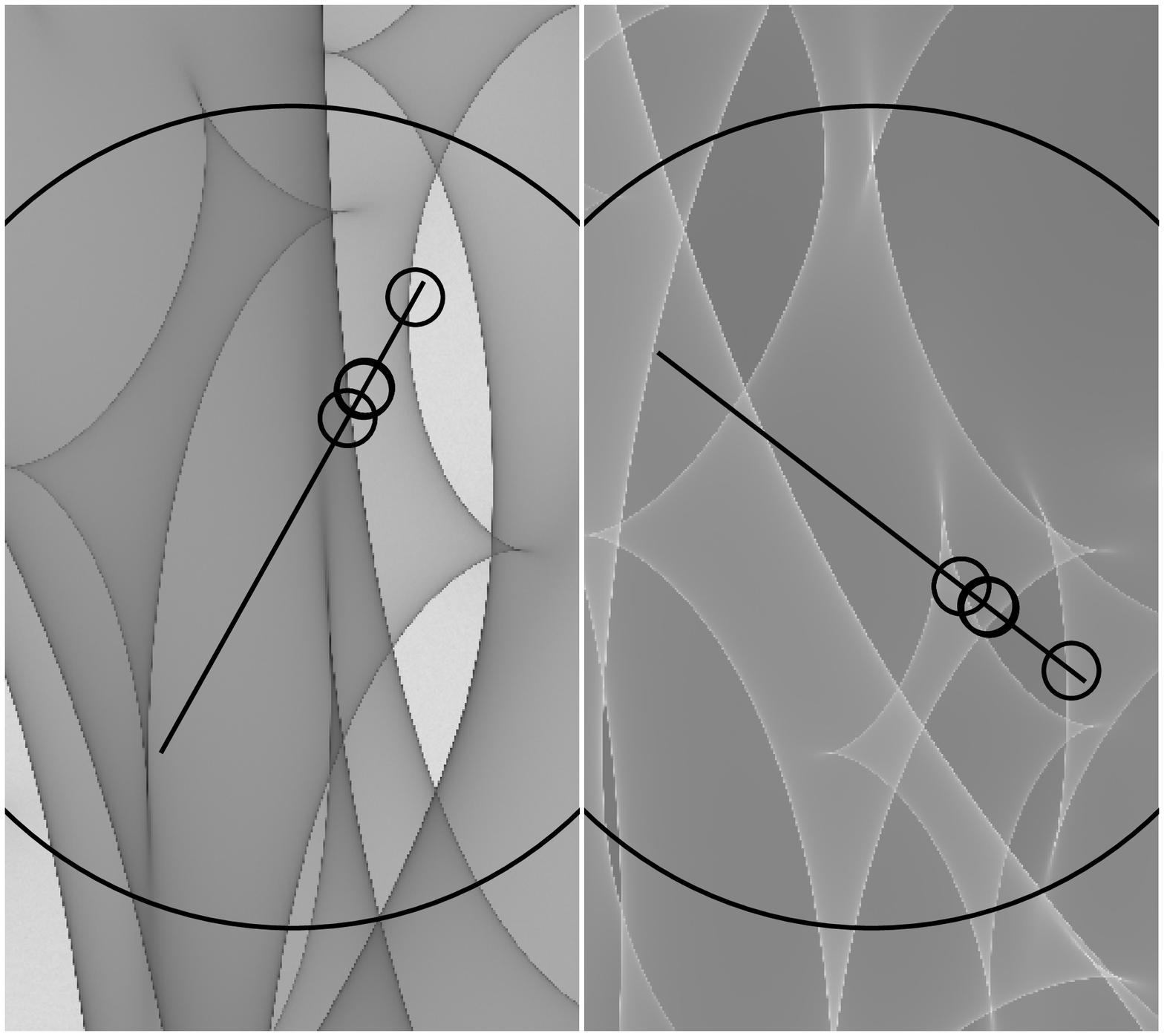,width=3.5in}}
\caption{The X-ray (top) and R-band optical (bottom) flux ratios between
the A$-$B and B$-$C images along with the tracks across the microlensing
patterns for images A (left) and B (right).    
The large circle shown on each pattern is the Einstein radius,
while the small circles have the half-light radius of the optical disk 
and are shown at the positions corresponding to the epochs of the X-ray 
observations.  The overall length of the line corresponds to one decade
of motion.  Darker colors represent logarithmically higher magnifications
with an overall magnification range from $1/30$ to $30$.  
This is a Case I example with fairly large differential
offsets.  It has a high stellar surface density ($f_*=0.7$), a large
amount of smooth optical emission ($f_{\hbox{no}\mu}=0.4$), and the X-ray source
is 14 times smaller than the optical.
}
\label{fig:lcurve1}
\end{figure}

The optical data consist of the five seasons of R-band 
monitoring data described in
Koz{\l}owski et al. (2009).  For our present analysis we simply shifted
the light curves by their measured time delays (Koz{\l}owski et al. 2009).
The X-ray data, all ACIS observations from the Chandra Observatory,
consist of the epoch presented by Blackburne et al. (2006) plus
the 5 epochs presented in Chartas et al. (2009).  Each of the
Chartas et al. (2009) epochs consisted
of a 5~ksec observation using ACIS-S3 in 1/8 sub-array mode 
from which we measure the 0.2-10~keV flux.  Chartas et al. (2009) also reanalyzed
the Blackburne et al. (2006) data to properly correct for 
the ``pile-up'' effect.  
We do not use the X-ray fluxes of image D in our analysis because
we cannot presently be certain its flux ratios relative to A--C
are unaffected by source variability given the roughly $3$ month 
time delay between D and A--C (Koz{\l}owski et al. 2009).  As we
can see from Fig.~\ref{fig:lcurve1}, the X-ray source must be more
compact than the optical source because the X-ray flux ratios are
dramatically more variable.

A full description of our microlensing analysis method is presented in Kochanek (2004)
and Kochanek et al. (2006).  In essence, we create the microlensing
magnification patterns we would see for a broad range of lens models
and source sizes, then randomly generate light curves to find ones
that fit the data well.  We then use Bayes' theorem to combine the
results for the individual trials to infer probability distributions
for physically interesting variables including the uncertainties 
created by all the other variables.  

We fit the lens as in Koz{\l}owski et al. (2009), modeling it as a 
$R_e=1\farcs7$ de Vaucouleurs
model for the stellar distribution embedded in an NFW halo.  We consider
a sequence of models described by $f_*$, the fraction of mass in the
stellar component relative to a constant mass-to-light ratio model with
$f_*\equiv 1$ and no halo. We include models with $f_*=0.1$ to $1$ in equal
steps, and the time delay measurements favor $f_* \simeq 0.2$.  These
lead to the values for the convergence $\kappa$, shear $\gamma$ and fraction of the
convergence in stars $\kappa_*/\kappa$ reported in Table.~1.

The stars creating the microlensing magnification were drawn from a power 
law mass function $dN/dM \propto M^{-1.3}$ with a ratio of 50 between the 
minimum and maximum masses that roughly matches the Galactic disk mass
function of Gould (2000).  We know from previous theoretical studies
that the choice of the mass function will have little effect on
our conclusions given the other sources of uncertainty (e.g. Paczy\'nski 1986,
Wyithe et al. 2000a). 
 The mean mass $\langle M\rangle$ is left as a 
variable with a uniform prior over the mass range $0.1 < \avgmhat < 1.0$.

 For each model we generated 8 random
realizations of the star fields near each image.
The magnification patterns had an outer scale of
$10\langle R_E\rangle = 4.6 \avgmhat^{1/2} \times 10^{17}$~cm
and a pixel scale of $10 \langle R_E\rangle/8192=5.6 \avgmhat^{1/2} \times 10^{13}~\hbox{cm}\simeq 3 r_g$,
so we should be able to model sources as compact as the inner edge of
the accretion disk.
We modeled the relative velocities as in Kochanek (2004), 
where for RXJ1131 the projection of the CMB dipole velocity (Kogut et al. 1993) on the
lens plane is 47~km/s, the lens velocity dispersion estimated
from the Einstein radius is
$350$~km/s, and the estimated rms peculiar velocities
of the lens and source galaxies are $180$ and $140$~km/s
respectively.  

The source model for both the optical and X-ray sources is a
face-on disk with a temperature profile $T\propto R^{-3/4}$
radiating as a black body (Shakura \& Sunyaev 1973), so the 
surface brightness profile of the disk is
\begin{equation}
     I(R) \propto \left[\exp( (R/R_\lambda)^{3/4}) -1 \right]^{-1}
\end{equation}
with the single parameter being the scale length $R_\lambda$.  While
it is true that this profile lacks the central drop in emissivity
and that it is not a physical model for the non-thermal X-ray emission, the
microlensing analysis is not sensitive to these details.  
The estimate of the half-light radius ($R_{1/2} \simeq 2.44 R_\lambda$)
is essentially independent of the assumed profile
(Kochanek 2004, Mortonsen, Schechter \& Wambsganss 2005). We used
a $46 \times 61$ logarithmic grid of trial source sizes for the 
X-ray and optical sources with a spacing of $0.05$~dex.

We do, however, allow for the possibility that fraction $f_{\hbox{no}\mu}=0$ 
to $40\%$ of the optical emission is generated on scales
much larger than the disk and is unaffected by microlensing.  
Such large scale emission could have two physical origins.
First, the optical continuum can be significantly contaminated by 
emission lines, both the obvious broad lines
and the less obvious Fe and Balmer pseudo-continuum emission ($\sim 30\%$
of the emission in some Seyferts, Maoz et al. 1993), that
are believed to be produced on much larger scales than the disk.
For our R-band light curves, there are no strong emission lines in
the filter band pass, but the blue edge of the Balmer continuum
emission ($\sim 6000$\AA) does lie inside the band pass (roughly
$5700$--$7200$\AA). Second, even if the observed photons were
generated by the accretion disk, a fraction could be scattered
on much larger scales, leading to an effectively larger source.
These two possibilities are not equivalent, as the line emission
is due to reprocessing of shorter wavelength UV photons rather
than the observed R-band continuum. 

A basic problem for any microlensing analysis is the degree to
which the ``macro'' lens model correctly sets the average magnifications.
Each light curve, $m_i(t) = s(t) + \mu_i + \delta\mu_i(t) +\Delta_i$
is defined by the source light curve $s(t)$, the macro model magnification 
$\mu_i$, the microlensing magnification $\delta \mu_i(t)$ and a possible 
offset $\Delta_i$.  These offsets can be non-zero
due to problems in the macro model or the presence of unrecognized substructures
that perturb the magnifications (e.g. Kochanek \& Dalal 2004), because of differential
absorption due to dust or gas in the lens galaxy (e.g. Falco et al. 1999, Dai
\& Kochanek 2009), or due to
contamination of the light curves by flux from the quasar host
or lens galaxy.   For the latter two possibilities, the offsets
would differ between the optical and X-ray light curves.  Given
a sufficiently long light curve, the offsets can be determined
from the data, but they are poorly constrained until the light
curve is a good statistical sampling of the magnification pattern.
We will consider four treatments of this problem to ensure that
such systematic problems do not affect our results.  The
basic division we will refer to as Cases I and II.  In Case I
we allow the magnification offsets $\Delta_i$ to float independently 
for the two bands constrained by a term $\Delta_i^2/2\sigma^2$ 
in the log likelihood
with $\sigma=0.5$~mag.  In Case II we allow them
to float, but use the same offsets $\Delta_i$ for both the optical and
X-ray light curves.  These are weak constraints, so the 
resulting distributions for the offsets are broad.  To make
sure we are not allowing too much freedom, we also examined
limiting the range of the offsets to $| \Delta_i| < 0.3$~mag
in Cases I' and II'.
 
The advantage of the less constrained strategies is that they
are robust 
against the systematic errors that can plague the absolute magnifications
of the images.  It is certainly true that analyses using only the
``DC'' flux ratios (e.g. Pooley et al. 2006, 2007, Bate et al. 2009, 
Floyd et al. 2009) require less data than our ``AC'' approach,
but they can also lead to conclusions dominated by these systematic errors.  
The ``AC'' approach
also has the advantage that including the effects of the velocities
allows us to estimate source sizes in centimeters 
without simply assuming a mean mass $\langle M\rangle$.
However, when we use loose priors on the DC flux ratios, we 
lose significant information on the locations of the images relative to
the magnified and demagnified regions of the patterns.  As such, it
is a conservative approach.  We consider all four offset treatments
in order
to explore their consequences on estimates of the source size and
the amount of dark matter in the lens.

We used 8 statistical realizations of the microlensing magnification 
patterns for each of the
10 stellar surface densities ($f_*$) and for 5 un-microlensed fractions 
of optical light ($f_{\hbox{no}\mu}$).  We modeled the data sequentially,
making $10^6$ trials for each optical source size and case, and then
fitting each trial that was a reasonable statistical fit to the optical data
to the X-ray data for each of the X-ray source sizes.   In this second
step we considered both the case where the X-ray and optical share
the same intrinsic flux ratios and where they are allowed to differ.

\begin{figure}[t]
\centerline{\psfig{file=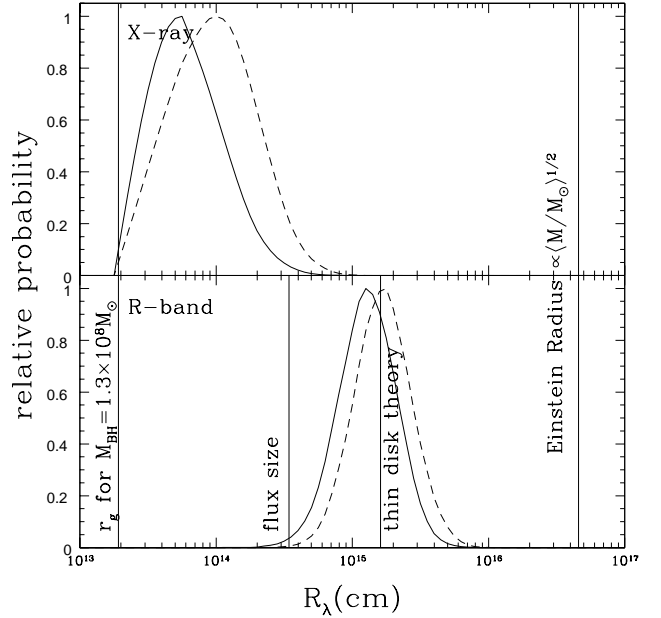,width=3.4in}}
\caption{The probability distributions for the size of the X-ray (top) and R-band
($400$~nm in the rest frame) optical (bottom) emission regions for the log (solid)
and linear (dashed) size priors.  These sizes are marginalized over $f_{\hbox{no}\mu}$.
The vertical lines mark the gravitational radius $r_g$ for a $M_{1131}$ black hole,
the Einstein radius for $\langle M \rangle = M_\odot$ and the accretion disk
size estimates based on either the I-band flux (Eqn.~\ref{eqn:fsize}) or
thin disk theory (Eqn.~\ref{eqn:rdisk}).  The microlensing
sizes and the I-band flux estimates can also be scaled by a $(\cos i)^{-1/2}$
inclination dependence from the assumed face-on case ($i=0^\circ$).
 }
\label{fig:dsize}
\end{figure}

\begin{figure}[t]
\centerline{\psfig{file=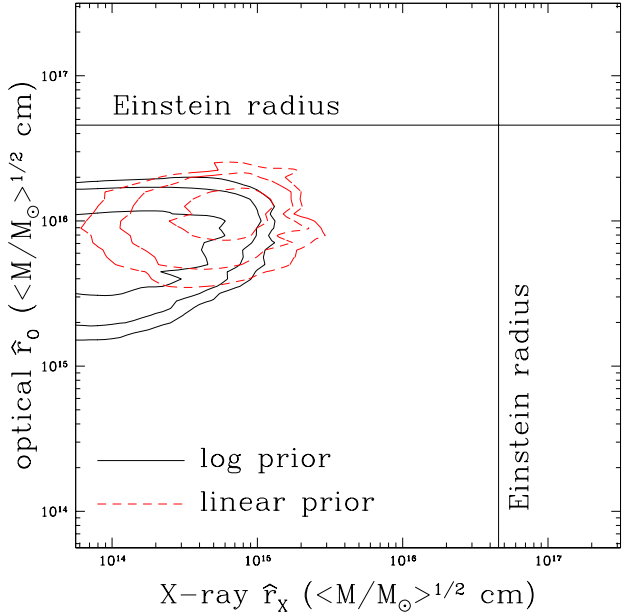,width=3.4in}}
\caption{The correlated probability distributions for the size of the optical and
  X-ray source sizes in Einstein units of $\langle M/M_\odot\rangle^{1/2}$~cm. 
  The contours are drawn at the 68\%, 90\% 
  and 95\% maximumum likelihood contours for one variable for log (solid)
  and linear (dashed) size priors.
 }
\label{fig:sizerat}
\end{figure}

\begin{figure}[t]
\centerline{\psfig{file=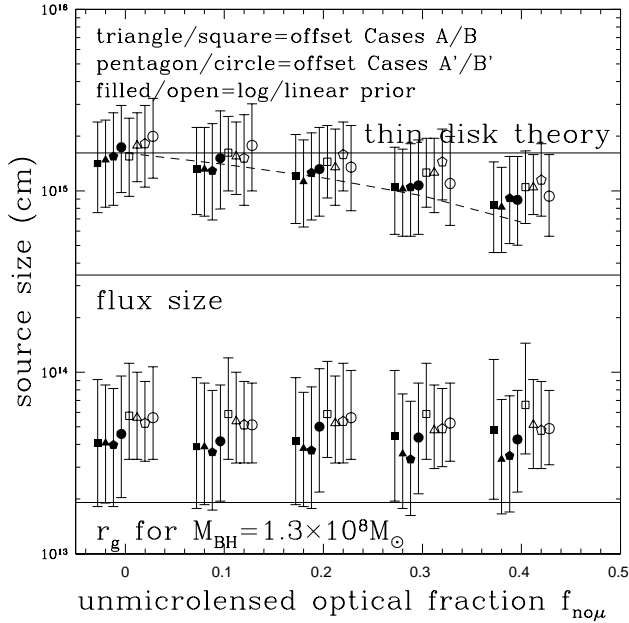,width=3.4in}}
\caption{Source size dependence on parameters.  The optical (top)
  and X-ray (bottom) source sizes ($R_\lambda$) as a function of
  the fraction $f_{\hbox{no}\mu}$ of the optical flux that is not microlensed.
  The triangles, squares, pentagons and circles show the results for the
  Case I (independent magnitude offsets for both bands), Case II (common 
  magnitude offsets), Case I' (independent offsets limited to $|\Delta_i|<0.3$)
  and Case II' (common  offsets limited to $|\Delta_i|<0.3$)
  treatments of the magnitude offsets.  The filled (open) symbols
  show the results for the logarithmic (linear) priors on the source
  sizes.  The horizontal lines show the same physical scales as in
  Fig.~\ref{fig:dsize} and the dashed curve shows the expected scaling
  of the optical size with $f_{\hbox{no}\mu}$ if we keep the half-light radius
  of the optical emission fixed.  The half light radius of the disk emission
  is always $R_{1/2}=2.44R_\lambda$, but the half light radius of the disk emission
  combined with the unmicrolensed large scale emission grows with $f_{\hbox{no}\mu}$,
  reaching $R_{1/2}=5.87 R_\lambda$ for $f_{\hbox{no}\mu}=0.4$. 
 }
\label{fig:cases}
\end{figure}

\begin{figure}[t]
\centerline{\psfig{file=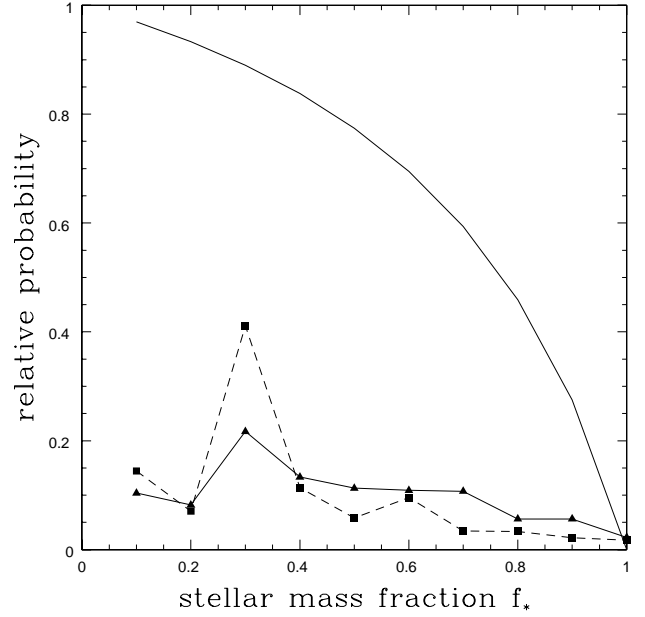,width=3.4in}}
\caption{Dependence on halo structure.  The solid/squares and dotted/triangles
  show the likelihood of $f_*$, the fraction of mass
  in the stellar component in the lens model compared to a constant $M/L$
  model ($f_*\equiv 1$), for weakly constrained (Case I+II) or strongly
  constrained (Case I'+II') treatments of the magnification offsets. 
  Dark matter dominated models
  are always favored, but the low $f_*$ models implied by the time delays
  are only strongly favored when we force the offsets to be small.
  The line without
  points shows the fraction $1-\kappa_*/\kappa$ of the local surface density near image A
  that is comprised of smoothly distributed dark matter.
 }
\label{fig:fstar}
\end{figure}

\begin{figure}[t]
\centerline{\psfig{file=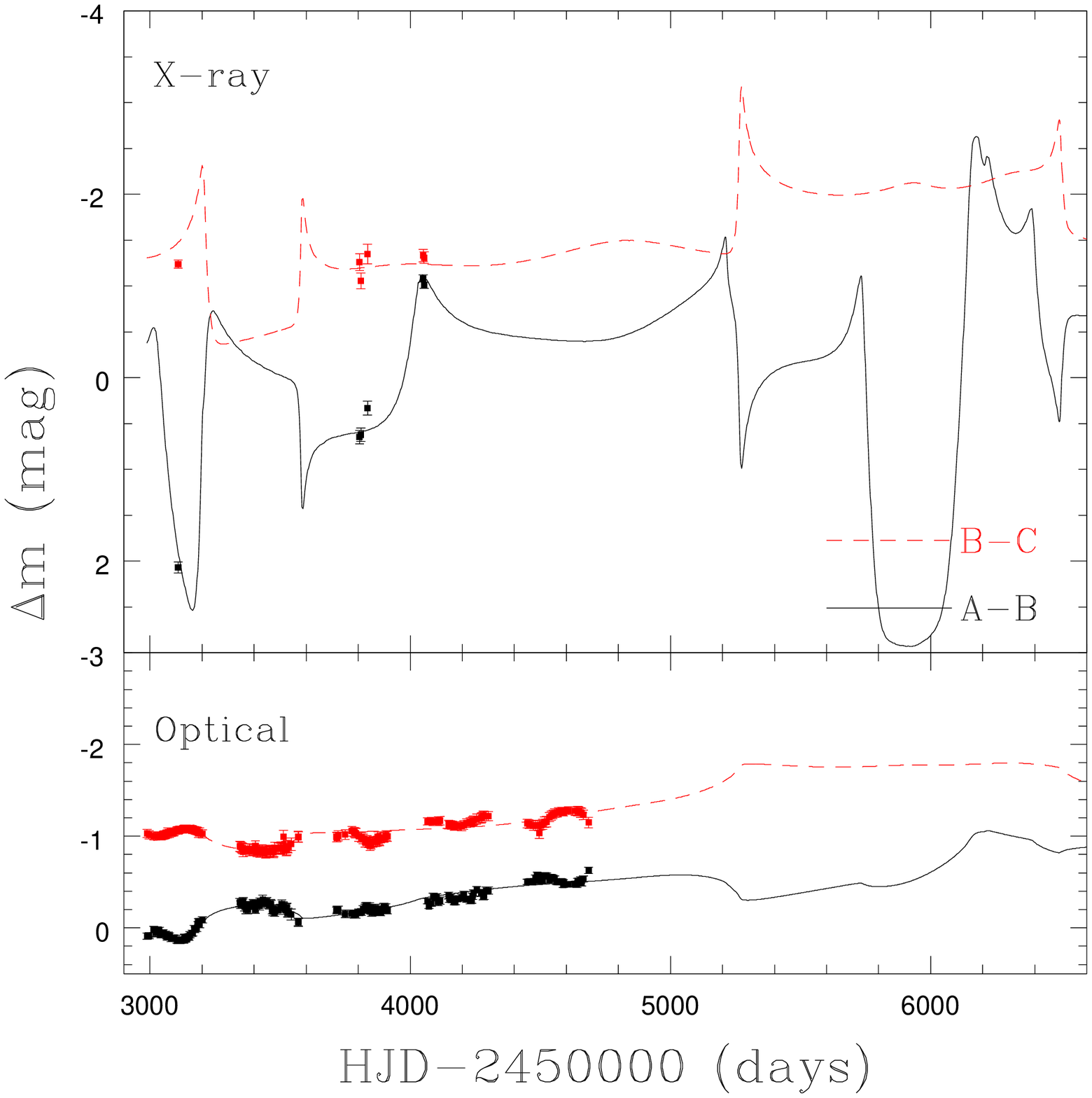,width=3.5in}}
\centerline{\psfig{file=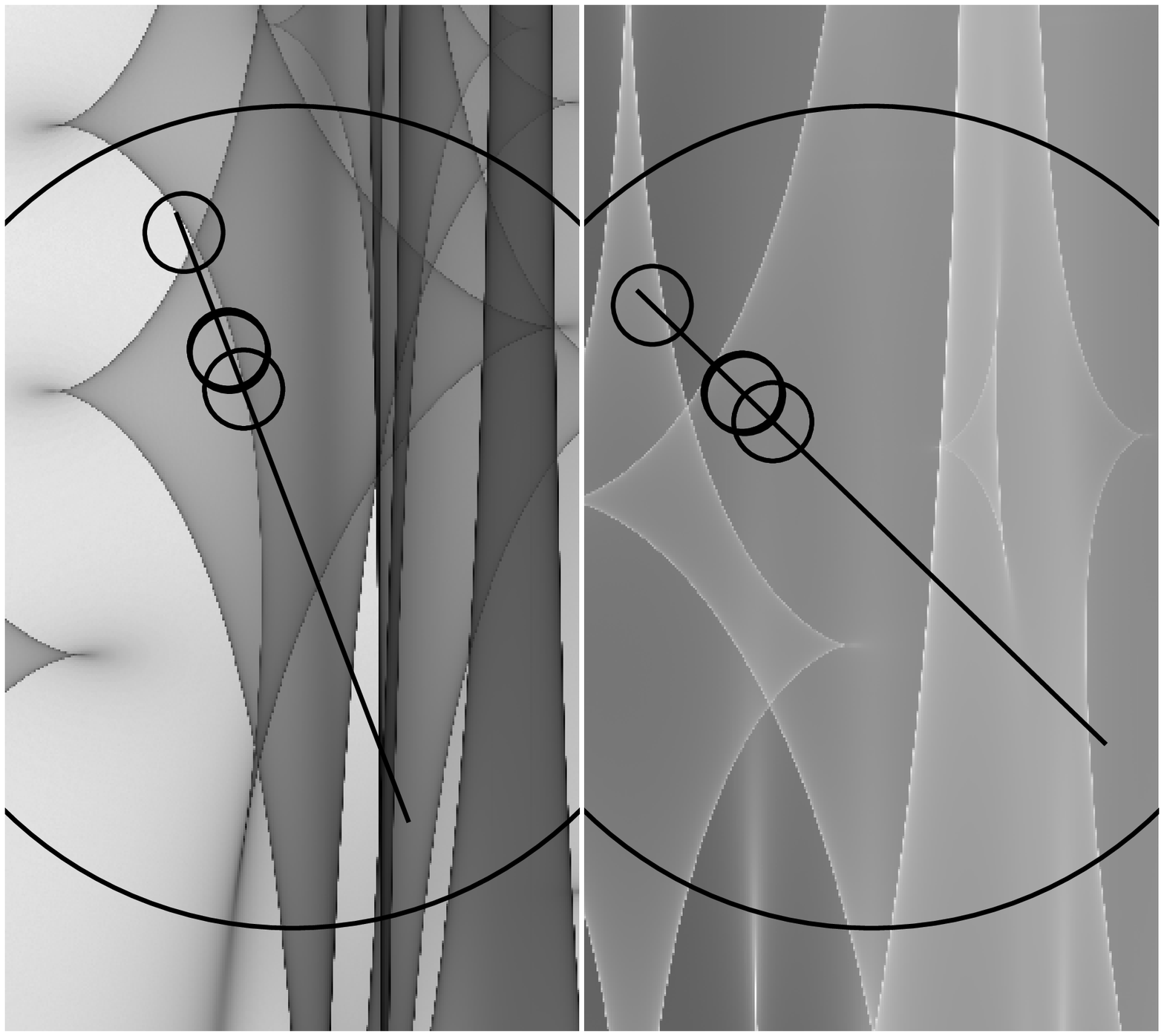,width=3.5in}}
\caption{ As in Fig.~\ref{fig:lcurve1}.  This is a Case II' example,
so the magnification offsets are small and the same for both the
optical and X-ray data.  It has a very low stellar surface density ($f_*=0.1$), 
a little smooth optical emission ($f_{\hbox{no}\mu}=0.2$), and the X-ray source
is 32 times smaller than the optical.
}
\label{fig:lcurve2}
\end{figure}

\begin{figure}[t]
\centerline{\psfig{file=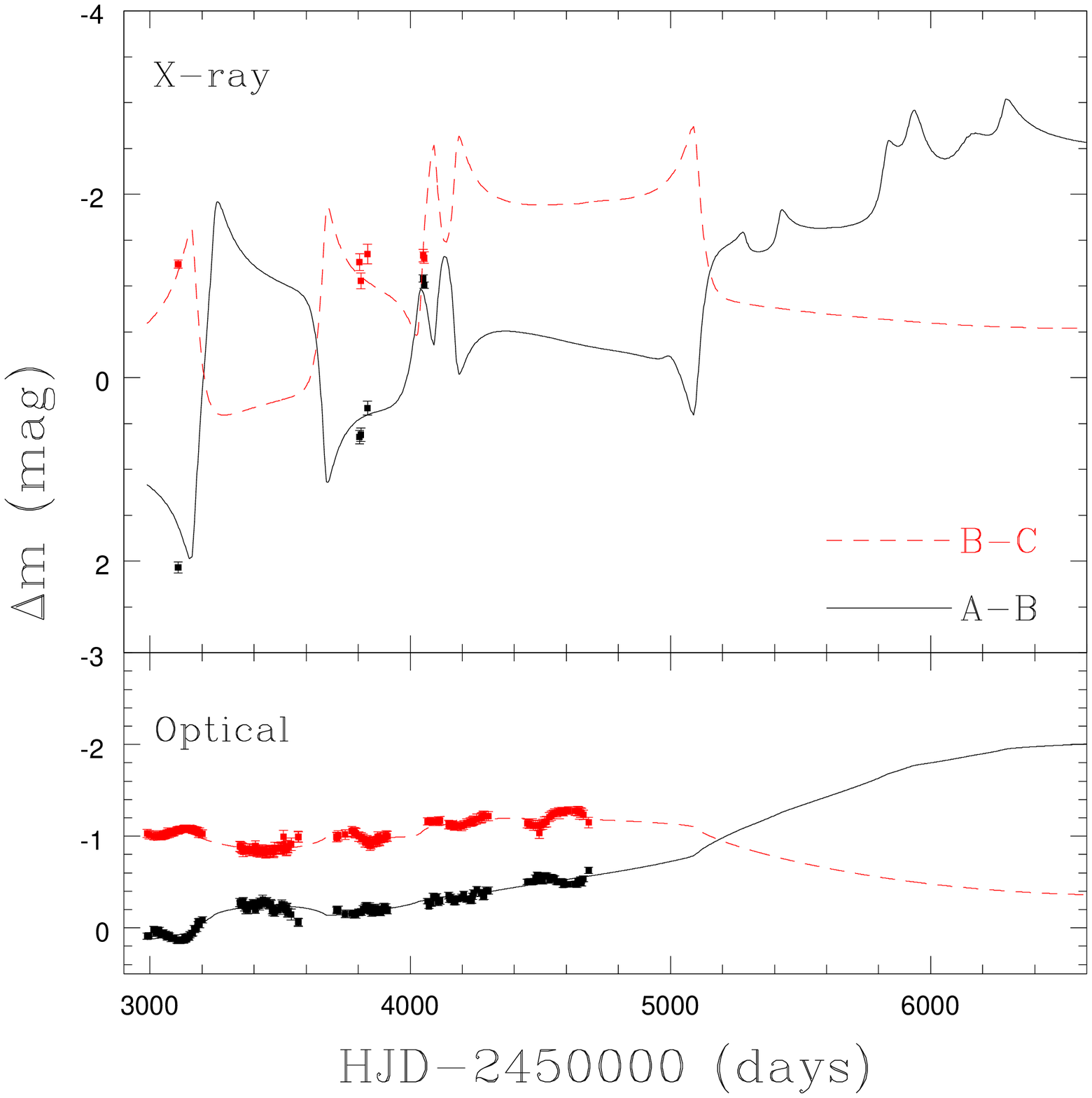,width=3.5in}}
\centerline{\psfig{file=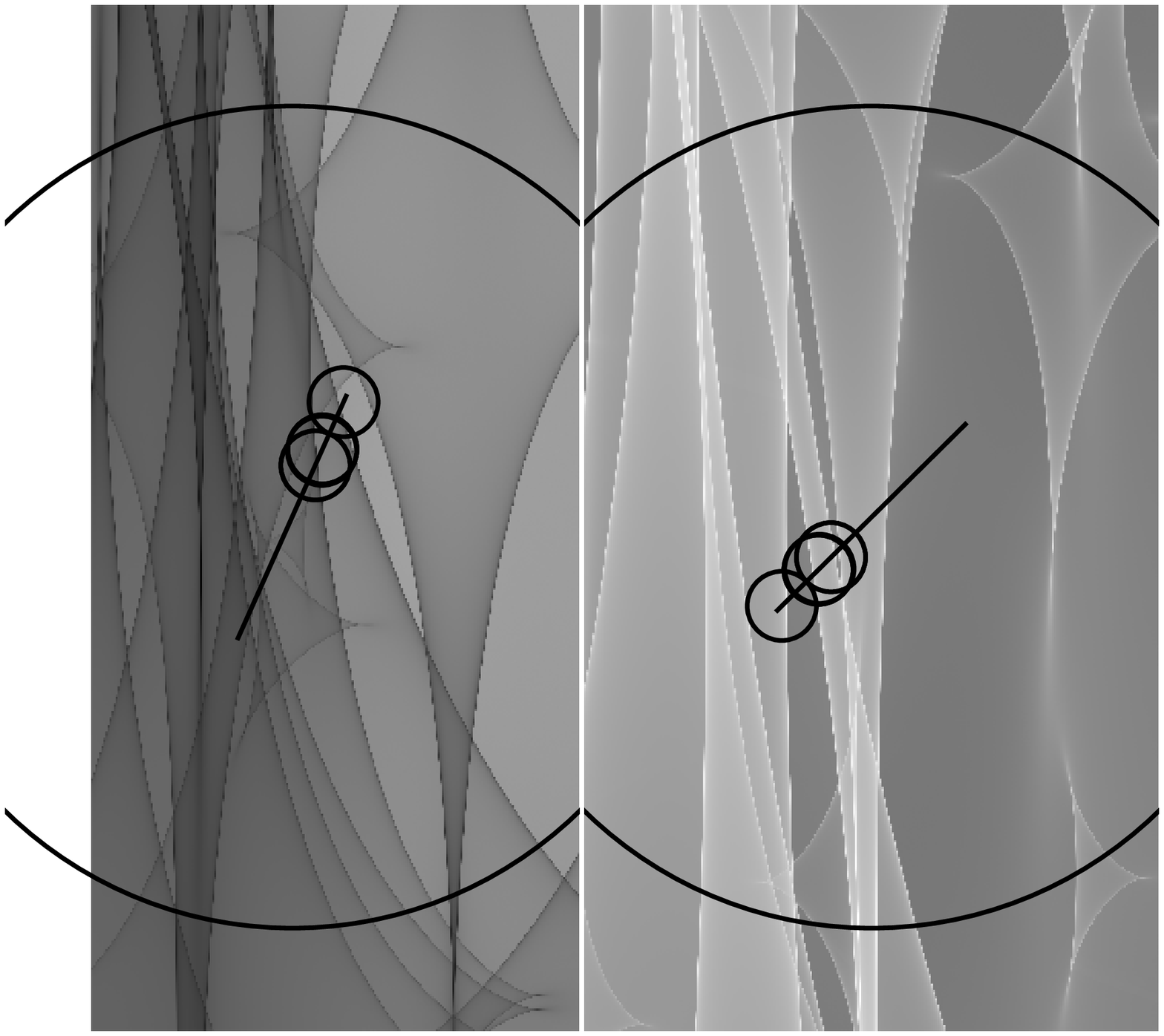,width=3.5in}}
\caption{ As in Fig.~\ref{fig:lcurve1}.  This is a Case I' example,
so the magnification offsets are small but differ between the
optical and X-ray data.  It has a low stellar surface density ($f_*=0.3$),
a little smooth optical emission ($f_{\hbox{no}\mu}=0.2$), and the X-ray source
is 28 times smaller than the optical.  In the left panel we are at the
edge of the pattern (although the Kochanek (2004) pattern creation method
here produces periodic patterns that allow
us to wrap the light curves across edges).
}
\label{fig:lcurve3}
\end{figure}

\section{Results and Discussion \label{sec:results} }

Fig.~\ref{fig:dsize} shows the main result for the estimated
size of the X-ray and optical emission regions.  These combine
all four treatments of the magnification offsets.  Also note
that in order to preserve the meaning of size ratios in Fig.~\ref{fig:dsize}, we used
the scale $R_\lambda$ of a face-on disk for both.  Physically, the
X-ray emission is better characterized by its half light
radius, $R_{1/2}=2.44 R_\lambda$.  The scale length of the
thin disk also scales as $\cos^{-1/2} i$ if not viewed face on.   
We show the results for two different priors on the disk
sizes, a logarithmic ($P(R_\lambda) \propto 1/R_\lambda$) and a uniform ($P(R_\lambda) \propto \hbox{constant}$)
prior, and this has minor effects for the optical estimate and
significant effects for the X-ray estimate.  For the logarithmic
prior we formally find that the (face on) optical disk scale length is
$\log (R_{\lambda,O}/\hbox{cm}) = 15.11$ ($14.89 < \log (R_{\lambda,O}/\hbox{cm}) < 15.32$)
and that the X-ray half-light radius is 
$\log (R_{1/2,X}/\hbox{cm}) = 14.36$ ($14.04 < \log (R_{1/2,X}/\hbox{cm}) < 14.68$).
These estimates use both the prior on the velocities and a
uniform prior for the mass over the range $0.1 < \langle M/M_\odot\rangle <1$.
We will focus on results including this mass prior, but note that
if we make no assumption about $\langle M \rangle$, the sizes
change little.  With only the velocity priors we find
$\log (R_{\lambda,O}/\hbox{cm}) = 15.02$ ($14.75 < \log (R_{\lambda,O}/\hbox{cm}) < 15.27$)
and 
$\log (R_{1/2,X}/\hbox{cm}) = 14.02$ ($13.67 < \log (R_{1/2,X}/\hbox{cm}) < 14.38$).
The source sizes become a little bit smaller, but the net effect is
very modest for the reason outlined in Kochanek (2004).\footnote{In Einstein
units, one can achieve the observed variability using either a large
source moving rapidly or a small source moving slowly, with a 
degeneracy of roughly $\hat{r} \propto \hat{v}$.  We always 
impose a prior on the physical velocity 
$v \propto \hat{v} \langle M\rangle^{1/2}$,
so the physical source size 
$r \propto \hat{r} \langle M\rangle^{1/2} \propto \hat{v} \langle M\rangle^{1/2} \propto v$
is essentially independent of $\langle M \rangle$ given a prior on the velocity.  }

The X-ray size is more sensitive to the priors because the convergence
of the probability distributions for small sources is poor
when the light curve is sparsely sampled.  Fig.~\ref{fig:sizerat}
shows likelihoods for the source size in the Einstein units used
for the basic calculations, and we see that they converge for small
X-ray sizes when we use a linear prior but not for a logarithmic
prior.  The problem is not due to the pixel scale of the maps, but
due to the lack of a well-sampled peak in the X-ray data.  Very
small source sizes are constrained by the magnification peaks 
observed during a caustic crossing.  If the light curve only
samples up to some minimum physical distance from a caustic 
crossing, then it will constrain sources sizes significantly
smaller than that distance poorly and the likelihood function
will flatten for small source sizes.   A logarithmic size
prior then favors these small scales compared to a linear
prior, leading to significant differences.  Thus, our lower
limits on the size of the X-ray emission are at a minimum
prior dependent.  More conservatively, the results could be
interpreted as providing only an upper bound on the size of
the X-ray emitting region.
  
Fig.~\ref{fig:cases} shows how the sizes depend on the priors,
the treatment of the magnification offsets and the
fraction $f_{\hbox{no}\mu}$ of the optical emission 
that is unaffected by microlensing.  The X-ray size is 
affected only by the choice of the size prior.  The optical size is
only affected by $f_{\hbox{no}\mu}$.  There are no significant differences
between the results for the four magnification offset cases.  In
order to produce the same optical variability with a larger
fraction $f_{\hbox{no}\mu}$,
we must shrink the disk scale length $R_\lambda$. Mortonson
et al. (2005) argue that the effects of microlensing are largely
determined by the half-light radius of the source, which is
$R_{1/2}=2.44R_\lambda$ in the limit that $f_{\hbox{no}\mu}=0$.  As
we increase $f_{\hbox{no}\mu}$, the disk scale length shrinks roughly
by the amount needed to keep the half light radius constant,
with $R_{1/2}=5.87 R_\lambda$ when $f_{\hbox{no}\mu}=0.4$.  Note, 
however, that the scaling for this particular model
will break down when $f_{\hbox{no}\mu}=1/2$.

The larger values of $f_{\hbox{no}\mu}$ are
favored, with likelihood ratios of $0.35$, $0.40$, $0.64$,
$0.92$ and $1.0$ for $f_{\hbox{no}\mu}=0$, $0.1$, $0.2$, $0.3$ and
$0.4$ respectively.  
These differences are only marginally
significant, but they are in the sense of favoring (effectively)
a flatter temperature profile.  A flatter temperature
profile can help to reconcile the differences between the
larger microlensing and thin disk theory sizes as compared
to the smaller flux sizes.  Such flatter temperature profiles
are generally consistent with the studies of the optical/infrared
wavelength dependence of microlensing (Anguita et al. 2008, Eigenbrod et al. 2008,
Poindexter et al. 2008, Mosquera et al. 2009, Bate et al. 2009),
but are not required.  The one exception is Floyd et al. (2009),
who find a limit requiring a steeper temperature profile.
Some of this information is also present in the overall spectral
energy distribution, and it is a long standing problem that the
spectra of quasars do not match the predictions of thin disk
theory (see Koratkar \& Blaes 1999, Gaskell 2008).  

Whether increasing $f_{\hbox{no}\mu}$ helps to resolve the size discrepancies 
depends on the physical model for the contamination. Line emission is
reprocessed shorter wavelength emission, so as we increase
$f_{\hbox{no}\mu}$ we are also reducing the fraction of the observed 
emission due to the disk and the flux size also shrinks 
as $(1-f_{\hbox{no}\mu})^{1/2}$.   If, however, we view it
as scattering fraction $f_{\hbox{no}\mu}$ of the continuum emission
on some large scale, then the flux size estimate is unchanged
and the effect helps to reduce the discrepancy. 
Resolving the
discrepancy with $f_{\hbox{no}\mu}$ would require that most of the
optical emission does not reach us directly from the accretion disk. 

While the source sizes show little dependence on the treatment
of the magnification offsets, estimates of the amount of dark
matter in the lens are sensitive to how strongly we 
constrain the models to match the observed macro model flux
ratios, as illustrated in Fig.~\ref{fig:fstar}.  By leaving
the offsets relatively free, so as to conservatively estimate
the source sizes, we have not optimized the calculation for
probing dark matter.  The Case I and II models, where we very loosely 
constrain the allowed magnification offsets, marginally favor
models with $f_* \simeq 0.3$.  The Case I' and II' models,
where we only accept small offsets, favor the same 
dark matter dominated model more strongly. This range for
$f_*$ is also that favored by the time delays measurements from 
Koz{\l}owski et al. (2009).  Note, however, that we are
not in the ``lagoon and caldera'' regime for the
microlensing patterns noted by Schechter \& Wambsganss (2002)
because of the very high magnifications ($\mu \sim 50$ for
image A at low $\kappa_*/\kappa$ rather than $\mu \sim 10$).

Figs.~\ref{fig:lcurve1}-\ref{fig:lcurve3} show several examples of model 
light curves that fit the data reasonably well.  These were also selected to 
have velocities consistent with masses of order $0.1$-$1.0M_\odot$.
The solutions are not unique, but they illustrate how simple 
changes in the source size dramatically alter the amplitude of the 
variability.  A common theme to the solutions is that the A and B
images are generally required to lie in ``active'' regions of the
patterns in order to produce such large changes in the X-ray fluxes.
This means that we can expect the dramatic variability observed in
this system to continue for an extended period of time ($1$--$10$~years).  
It is
also interesting to note that significant changes in the optical
fluxes are also likely.  The larger optical source size washes
out the effects of the closely spaced caustics that help to 
drive the X-ray variability.  But the overall changes between
the high magnification ridges and the demagnified valleys are 
still significant, and we should see overall changes in the 
optical fluxes several times those observed to date.

The implications of these results for theoretical models are mixed.
The size of the disk is grossly similar to that expected from thin
disk theory, and as we have summarized in Morgan et al. (2009),
we find disk sizes that scale with black hole mass and optical
wavelength roughly as expected (Eqn. \ref{eqn:rdisk}).  We also
find that the X-ray emission arises from significantly closer
to the expected inner edge of the accretion disk than the 
optical emission, as we might expect for a hot corona. 
The optical size is broadly 
consistent with the expectations for an Eddington luminosity
black hole with a mass estimated from the emission line widths
(Eqn.~\ref{eqn:rdisk}).  But the size is inconsistent with 
that expected for a thermally radiating disk with the observed 
magnitude (Eqn.~\ref{eqn:fsize}). This is the discrepancy
originally noted by Pooley et al. (2006), which we explore
more quantitatively in Morgan et al. (2009).

Should we conclude that the thin disk model is wrong or simply
that we have oversimplified the optical radiation transfer?
We considered contamination by line emission or scattering
of the optical photons, finding that this can modestly reduce
the disk size for the range where up to 40\% of the optical
emission does not come directly from the disk.
Our simple emission model neglects the disk atmosphere and 
heating of the outer disk by radiation from the inner disk,
all processes which would tend to make the optical emission 
region larger than the point where the disk has a temperature
matching the photon wavelength without any change in the
underlying properties of the disk.  Many of these effects are
included in recent disk models such as Hubeny et al. (2001) or Li et al. (2005)
for disk spectra.  We examined face-on models with $M_{BH}=10^8M_\odot$,
$\dot{M}=0.09 M_\odot$~yr$^{-1}$ and a BH spin of $a=0.998$ using
the Hubeny et al. (2001) models to compute our definition of
the disk scale ($R_\lambda$, where $kT=hc/\lambda$) and the half-light radius
($R_{1/2}$).
The scale $R_\lambda$ is the most sensitive to the 
assumptions, with $R_\lambda/r_g=41$ for $\lambda_{obs}=0.814\mu$m
in our simplified disk model
but equal to $36$/$34$ for black body non-relativistic/relativistic
disk models (BB NR/REL) and to $28$/$26$ for non-LTE 
non-relativistic/relativistic models (NLTE NR/REL).  The
model dependence is much reduced if we compare with the
half-light radii ($R_{1/2}/r_g=100$ for the simple model, $114$/$117$
for the NLTE NR/REL models, and $99$/$102$ for the BB NR/REL
models).  A flatter temperature profile than $T \propto R^{-3/4}$
would help, since at fixed total flux the half light radius 
increases.  For example if $T\propto R^{-1/2}$, the flux would be only 
20\% that of our standard profile for the same half-light radius. 
Indeed, such a flat temperature profile would also come much closer
to matching the observed spectra of quasars (e.g. Koratkar \& Blaes 1999, 
Gaskell 2008)
and would be representative of models dominated by irradiation.
In general, however, current microlensing results on temperature
profiles do not favor such flat profiles even if they generally
allow somewhat flatter profiles (e.g. Anguita et al. 2008, 
Eigenbrod et al. 2008, Poindexter et al. 2008, Mosquera et al. 2009, 
Bate et al. 2009, Floyd et al. 2009).

The key to disentangling these problems is to expand the measurements
over a broad range of wavelengths, so that we can constrain the 
temperature profile of the disk, and over a broad range of 
black hole masses and accretion rates.  For the particular
case of RXJ1131, we have programs to continue the
X-ray monitoring of the system and to use
HST to monitor the ultraviolet flux ratios of the images.   
Obtaining a robust lower limit to the size of the X-ray emitting
region may require denser sampling of the X-ray light curve.
Measuring the mid-infrared flux ratios of the images would be
the most important step towards rigorously imposing the
relative macro magnifications with no offsets, as 
these would only be affected by the macro model and 
any large scale gravitational perturbations (satellites).

\acknowledgements 

We would like to thank M. Dietrich, P. Osmer, B. Peterson and R. Pogge for 
many discussions on quasar structure, and O. Blaes for his
comments.  We would like to thank D. Sluse for supplying their spectrum of the system.
The calculations in this paper were carried out on a Beowulf cluster obtained
as part of the Cluster Ohio program of the Ohio Supercomputer Center.  
Support for this work was provided by NASA through Chandra
Award GTO-07700072 and by the NSF through grant AST~0708082.

\clearpage
\begin{deluxetable}{lclll}
\tablewidth{0pt}
\tablecaption{Microlensing Model Parameters}
\tablehead{
   \colhead{$f_*$}         & \colhead{Image}   
  &\colhead{$\kappa$}      & \colhead{$\gamma$}
  &\colhead{$\kappa_*/\kappa$}
  }
\startdata
\input kapgam
\tableline
\enddata
\tablecomments{The macro models are parametrized by $f_*$, the fraction of mass in 
the de Vaucouleurs component relative to a constant $M/L$ model with $f_*\equiv 1$.
The microlensing model parameters are the convergence $\kappa$, shear $\gamma$ and 
the fraction of the convergence in stars $\kappa_*/\kappa$.}
\end{deluxetable}

\end{document}

%% file: kapgam.tex
0.1 &A & 0.667 & 0.359 & 0.030 \\ 
    &B & 0.631 & 0.325 & 0.027 \\ 
    &C & 0.644 & 0.306 & 0.028 \\ 
    &D & 1.079 & 0.493 & 0.079 \\ 
0.2 &A & 0.618 & 0.412 & 0.067 \\ 
    &B & 0.581 & 0.367 & 0.060 \\ 
    &C & 0.595 & 0.346 & 0.062 \\ 
    &D & 1.041 & 0.631 & 0.159 \\ 
0.3 &A & 0.569 & 0.465 & 0.110 \\ 
    &B & 0.530 & 0.410 & 0.099 \\ 
    &C & 0.546 & 0.387 & 0.103 \\ 
    &D & 1.001 & 0.635 & 0.242 \\ 
0.4 &A & 0.519 & 0.518 & 0.162 \\ 
    &B & 0.480 & 0.453 & 0.146 \\ 
    &C & 0.496 & 0.427 & 0.153 \\ 
    &D & 0.964 & 0.895 & 0.329 \\ 
0.5 &A & 0.469 & 0.572 & 0.226 \\ 
    &B & 0.430 & 0.497 & 0.204 \\ 
    &C & 0.447 & 0.469 & 0.214 \\ 
    &D & 0.925 & 1.018 & 0.421 \\ 
0.6 &A & 0.419 & 0.626 & 0.305 \\ 
    &B & 0.379 & 0.540 & 0.278 \\ 
    &C & 0.397 & 0.511 & 0.290 \\ 
    &D & 0.890 & 1.139 & 0.520 \\ 
0.7 &A & 0.369 & 0.679 & 0.406 \\ 
    &B & 0.329 & 0.584 & 0.375 \\ 
    &C & 0.348 & 0.552 & 0.390 \\ 
    &D & 0.851 & 1.288 & 0.625 \\ 
0.8 &A & 0.318 & 0.734 & 0.541 \\ 
    &B & 0.278 & 0.628 & 0.507 \\ 
    &C & 0.297 & 0.595 & 0.524 \\ 
    &D & 0.816 & 1.412 & 0.740 \\ 
0.9 &A & 0.268 & 0.787 & 0.725 \\ 
    &B & 0.228 & 0.671 & 0.697 \\ 
    &C & 0.247 & 0.637 & 0.711 \\ 
    &D & 0.781 & 1.530 & 0.863 \\ 
1.0 &A & 0.217 & 0.842 & 1.000 \\ 
    &B & 0.178 & 0.714 & 1.000 \\ 
    &C & 0.196 & 0.679 & 1.000 \\ 
    &D & 0.740 & 1.639 & 1.000 \\ 

%% file: ms.bbl
\begin{thebibliography}{}

\bibitem[Agol et al.(2009)]{2009ApJ...697.1010A} Agol, E., Gogarten, S.~M., 
Gorjian, V., \& Kimball, A.\ 2009, \apj, 697, 1010 
\bibitem[]{}Anguita, T., Schmidt, R.W., Turner, E.L., Wambsganss, J., Webster, R.L., Loomis, K.A.,
   Long, D., \& McMillan, R., 2008, A\&A, 480, 327
\bibitem[]{}Bate, N.F., Floyd, D.J.E., Webster, R.L., \& Wyithe, J.S.B., 2008, MNRAS, 391, 1955
\bibitem[]{}Bentz, M.C., Peterson, B.M., Pogge, R.W., Vestergaarrd, M., Onken, C.A., 2006, ApJ, 644, 133
\bibitem[]{}Blackburne, J.A., Pooley, D., \& Rappaport, S., 2006, 
   ApJ, 640, 569
\bibitem[]{}Blaes, O.M., 2004, in Les Houches Summer School LXXVIII (Springer: Berlin) 137
\bibitem[]{}Chartas, G., Kochanek, C.S., Dai, X., Poindexter, S., \& Garmire, G., 2009, ApJ, 693, 174
\bibitem[]{}Dai, X., Chartas, G., Agol, E., Bautz, M.W., \& Garmire,
  G.P., 2003, ApJ, 589, 100
\bibitem[]{}Dai, X., \& Kochanek, C.S., 2009, ApJ, 692, 677
\bibitem[]{}Eigenbrod, A., Courbin, F., Meylan, G., Agol, E., Anguita, T., Schmidt, R.W., \& Wambsganss, J., 2008, A\&A, 490, 933
\bibitem[]{}Fabian, A.C., Miniutti, G., Iwasawa, K., \& Ross, R.R., 2005, MNRAS, 361, 795
\bibitem[]{}Falco, E.E., Impey, C.D., Kochanek, C.S., Leh\'ar, J.,
   McLeod, B.A., Rix, H.-W., Keeton, C.R., Mu\~noz, J.A. \&
   Peng, C.Y., 1999, ApJ, 523, 617 
\bibitem[]{}Floyd, D.J., Bate, N.F., \& Webster, R.L., 2009, MNRAS in press [arXiv:0905.2651]
\bibitem[]{}Gaskell, C.M., 2008, Revista Mexicana de Astronomia Y Astrofisica Conference Series, 32, 1 [arXiv:0711.2113]
\bibitem[]{}Goicoechea, L.J., Alcaide, D., Mediavilla, E., Mun\~oz, J.A., 2003, A\&A, 397, 517
\bibitem[Gould(2000)]{2000ApJ...535..928G} Gould, A.\ 2000, \apj, 535, 928 
\bibitem[]{}Hubeny, I., Blaes, O., Krolik, J.H., \& Agol, E., 2001, ApJ, 559, 680
\bibitem[]{}Kaspi, S., Maoz, D., Netzer, H., Peterson, B.M., Vestergaard, M., \& Jannuzi, B.T., 2006, ApJ, 629, 61
\bibitem[]{}Kochanek C.S., 2004, ApJ, 605, 58
\bibitem[]{}Kochanek, C.S., \& Dalal, N., 2004, ApJ, 610, 69
\bibitem[]{}Kochanek, C.S., Dai, X., Morgan, C., Morgan, N., Poindexter, S., \& Chartas, G.,
  2006, Statistical Challenges in Modern Astronomy IV, 2007, G. J. Babu and E. D. Feigelson (eds.), San Francisco:Astron. Soc. Pacific
  [astro-ph/0609112]
\bibitem[]{}Koratkar, A., \& Blaes, O., 1999, PASP, 111, 1
\bibitem[]{}Koz{\l}owski, S., Morgan, N.D., Kochanek, C.S., Poindexter, S., Falco, E.E., \& Dai, X., 2009, in preparation
\bibitem[]{}Kogut, A., et al., 1993, ApJ, 419, 1
\bibitem[]{}Li, L.-X., Zimmerman, E.R., Narayan, R., \& 
  McClintock, J.E., 2005, ApJS 157, 355 
\bibitem[]{}Maoz, D., Netzer, H., Peterson, B.M., et al., 1993, ApJ, 404, 576
\bibitem[]{}Morgan, C.W., Kochanek, C.S., Morgan, N.D., \& Falco, E.E., 2009, ApJ submitted [arXiv:0707.0305]
\bibitem[Morgan et al.(2008)]{2008ApJ...689..755M} Morgan, C.~W., Kochanek, 
  C.~S., Dai, X., Morgan, N.~D., \& Falco, E.~E.\ 2008, \apj, 689, 755 
\bibitem[]{}Mortonson, M.J., Schechter, P.L., \& Wambsganss, J., 
   2005, ApJ, 628, 594
\bibitem[Mosquera et al.(2009)]{2009ApJ...691.1292M} Mosquera, A.~M., 
Mu{\~n}oz, J.~A., \& Mediavilla, E.\ 2009, \apj, 691, 1292 
\bibitem[]{}Novikov, I.D., \& Thorne, K.S., 1973, in Black Holes, ed. C. De Witt \& B. De Witt
  (New York: Gordan \& Breach) 343
\bibitem[]{}Paczy\'nski, B., 1986, ApJ, 301, 503
\bibitem[]{}Page, D.N., \& Thorne, K.S., 1974, ApJ, 191, 499
\bibitem[]{}Peng, C.Y., Impey, C.D., Rix, H.-W., Kochanek, C.S.,  
  Keeton, C.R., Falco, E.E., Leh\'ar, J., \& McLeod, B.A., 2006
  ApJ in press [astro-ph/0603248] 
\bibitem[]{}Poindexter, S, Morgan, N, \& Kochanek, C.S., 2008, ApJ, 673, 34
\bibitem[Pooley et al.(2006)]{2006ApJ...648...67P} Pooley, D., Blackburne, 
J.~A., Rappaport, S., Schechter, P.~L., \& Fong, W.-f.\ 2006, \apj, 648, 67 
\bibitem[Pooley et al.(2007)]{2007ApJ...661...19P} Pooley, D., Blackburne, 
J.~A., Rappaport, S., \& Schechter, P.~L.\ 2007, \apj, 661, 19 
\bibitem[]{}Rauch, K.P., \& Blandford, R.D., ApJ, 1991, ApJL, 381, 39L
\bibitem[]{}Reynolds, C.S., \& Nowak, M.A., 2003, PhR, 377, 389
\bibitem[]{}Schechter, P.L., et al., 2003, ApJ, 584, 657
\bibitem[]{}Schechter, P.L. \& Wambsganss, J., 2002, ApJ, 580, 685
\bibitem[]{}Shakura, N.I., \& Sunyaev, R.A., 1973, A\&A, 24, 337
\bibitem[]{}Sluse, D., Surdej, J., Claeskens, J.-F., Hutsemekers, D., Jean, C., 
  Courbin, F., Nakos, T., Billeres, M., \& Khmil, S.V., 2003, A\&A, 406, L43
\bibitem[]{}Sluse, D., Claeskens, J.-F., Altieri, B., Cabanac, R.A., Garcet, O.,
  Hutsemekers, D., Jean, C., Smette, A., \& Surdej, J., 2006, A\&A, 449, 539
\bibitem[]{}Wambsganss, J., 2006, in Gravitational Lensing: Strong
  Weak and Micro, Saas-Fee Advanced Course 33, G. Meylan,
  P. North, P. Jetzer, eds., (Springer: Berlin) 453
  [astro-ph/0604278]
\bibitem[]{}Wambsganss, J., Schmidt, R.W., Colley, W.N., Kundi\'c, T., Turner, E.L.,
   2000, A\&A, 362, L37
\bibitem[]{}Wambsganss, J., Schneider, P., \& Paczy\'nski, B., 1990, ApJ, 358, L33
\bibitem[]{}Wyithe, J.S.B., Webster, R.L., \& Turner, E.L.,
  2000a, MNRAS, 315, 51
\bibitem[]{}Wyithe, J.S.B., Webster, R.L., \& Turner, E.L.,
  2000b, MNRAS, 315, 62


\end{thebibliography}
